\documentclass[12pt,cites,graphicx]{article}
\usepackage{epsfig}
\title{Structural, electronic and optical properties of tetrahedral $Si_xGe_{47-x}:H_{60}$ nanocrystals: A Density Functional study}
\author{C. S. Garoufalis\\[3mm]
Department of Physics, University of Patras, GREECE, 26504,
Patras}
\begin{document}
\maketitle
\renewcommand{\thefootnote}{\fnsymbol{footnote}}
\noindent The structural, cohesive, electronic and optical
properties of mixed SiGe:H quantum dots are studied by Density
Functional Theory (DFT) calculations on a representative ensemble
of medium size nanoparticles of the form $Si_xGe_{47-x}:H_{60}$.
The calculations have been performed in the framework of the
hybrid non-local exchange-correlation functional of Becke, Lee,
Parr and Yang (B3LYP). Besides the ground state DFT/B3LYP values
we provide reliable result for the lowest spin and symmetry
allowed electronic transition based on Time Dependent DFT
(TDDFT/B3LYP) calculations. Our results show that the optical gap
depends not only on the relative concentrations of silicon,
germanium and hydrogen, but also on the relative position of the
silicon and germanium shells relative to the surface of the
nanocrystal. This is also true for the structural, cohesive and
electronic properties allowing for possible electronic and optical
gap engineering. Moreover, it is found that for the cases of
nanoparticles with pure Ge or Si core, the optical properties are
mainly determined by the Ge part of the nanoparticle, while
silicon seem to act as a passivant.
\section{Introduction}
\label{intro}

The possibility of tunable photoluminescence (PL) from silicon and
silicon-like (e.g. germanium) quantum dots (and nanowires), has
stimulated intensive research on this type of materials over the
last decade \cite{1,2,3,4,5,6,7,8,9,10,11,12}. Until recently,
silicon nanocrystals have practically "monopolized" the interest
of the researchers. A large portion of this type of work has been
devoted to understanding the visible photoluminescence of these
materials and its dependence on the diameter of the nanoparticles.
It is widely accepted and well established by now (see for
instance refs. 12-15) that the luminescence of oxygen-free Si
nanocrystals (of well defined diameter) is mainly due to quantum
confinement (QC) of the corresponding nanoparticles. This is also
true for Ge nanoparticles\cite{7,8}. It is known that the effect
of quantum confinement is even more pronounced for the case of Ge
nanoparticles. This can be easily understood by comparing the
electron and hole effective masses and dielectric constants of Si
and Ge. In particular, the smaller electron and hole effective
masses of Ge along with the larger dielectric constant (compared
to Si) result in a larger exciton Bohr radius for Ge.
Consequently, it might be expected that the effect of QC on the
optical properties of Ge nanoparticles will be more pronounced.
The PL properties of such nanocrystals (Si or Ge) are mainly
controlled by suitably regulating the size of the nanocrystals and
in many cases, their surface passivation. The possibility of
combining the advantages of Si (in the electronic properties) with
those of Ge (especially structural and mechanical properties)
appears to be a natural extension of scientific interest and an
intriguing and potentially promising field for the development of
optoelectronic nanodevices. It has been demonstrated by both
experimental observation\cite{13} and theoretical
calculations\cite{13,14} that the lattice mismatch of Si and Ge
has a significant effect on the electronic properties of
$Si_{1-x}Ge_x$ alloys. The induced strain affects mainly the tail
of the conduction band which results in an almost linear decrease
of the indirect band gap. In this sense, it may be expected that a
similar behavior may introduce interesting optical features in
$Si_xGe_y:H_z$ nanocrystals. Several of these issues have been
recently addressed by Ming Yu et. al.\cite{20} in the framework of
Density Functional Theory. In particular, they have performed
DFT/LDA molecular dynamics calculations on medium size $Si_xGe_y$
and $Si_xGe_y:H$ ($x+y=71$) mixed nanoparticles. Especially for
the case of hydrogenated nanocrystals they found that the
dependence of the single particle HOMO-LUMO gap on the relative
composition of the clusters exhibits many similarities with the
corresponding one of the bulk $Si_{1-x}Ge_x$ alloys. At this point
it should be noted that the structure of the specific nanocrystals
has been fully relaxed through a molecular dynamics procedure with
an initial temperature of 1000 K. As a result a large portion of
the strain induced by the Si/Ge mismatch (in the initial geometric
configurations) has been largely relieved. However, it would be
interesting to expand the investigation for the case of mixed
Si/Ge nanocrystal which have not undergone such an annealing
procedure. In this case, the aforementioned strain can not be
fully relieved since the individual atoms are only allowed to a
local relaxation around their original position (i.e they are not
allowed to diffuse through the shells). With this in mind, we have
examined the optical and electronic properties of mixed
nanocrystals of the form $Si_xGe_{47-x}:H_{60}$. We have studied
in detail the variation of the cohesive, electronic and optical
properties as a function of $x$. Moreover, we have examined the
dependence of these properties on the position of each atomic
species relative to the nanocrystal's surface.
\section{Technical details of the calculations}
\label{tech} All ground state calculations in this work are based
on Density Functional Theory (DFT), while all excited state
calculations are based on TDDFT. In both cases we employed the
nonlocal exchange-correlation functional of Becke, Lee, Yang and
Parr (B3LYP) \cite{18}. The accuracy of these calculations
(TDDFT/B3LYP) for the optical gap has been tested before by by
comparison with high level multireference second-order
perturbation theory (MR-MP2) calculations for the case of Si
nanocrystals\cite{5}. The size of the $Si_xGe_{47-x}:H_{60}$
nanocrystals considered here is approximately 10-12\AA. The
symmetry of the nanocrystals is $T_d$ and their geometries have
been fully optimized within this symmetry constrain using the
hybrid B3LYP functional. To preserve the $T_d$ symmetry, we
substituted shells of silicon (rather than isolated atoms) by
equivalent germanium shells. This choice introduces an additional
restriction on the variation of Si concentration. This procedure
imposes some constrains in the relaxation of the interatomic
forces. In particular, although bond lengths an angles are allowed
to relax, the atoms are not allowed to change their relative
position in the nanoparticle. As a result, migration from the
inner core to the surface (or vice versa) is not possible (such
migrations were both allowed an observed in the MD calculations of
Ming Yu et al\cite{20}). We have examined in detail most of the
structural (bond length distribution), cohesive (binding
energies), electronic (DOS, electronic gaps) and optical
properties as a function of the concentration $x$. The optical gap
is defined as the energy of the lowest spin and symmetry allowed
excitation calculated by the TDDFT/B3LYP method. Moreover, for the
same concentration $x$ we have considered alternative ways of
substitution of the shells of silicon atoms by germanium. The bulk
of our calculations were performed with the TURBOMOLE \cite{15}
suite of programs using Gaussian atomic orbital basis sets of
split valence [SV(P)]: [4s3p1d]/[2s] quality \cite{16}. Test
calculations with the larger TZVP basis set revealed only marginal
deviations from the corresponding SV(P) results.
\section{Results and  discussion}
\label{res}
\subsection{Structural and cohesive properties}
Representative geometries of $Si_xGe_{47-x}:H_{60}$ nanoclusters
are shown in figure 1 for various concentrations and
substitutions. The bonding characteristics of the various
structures can be easily visualized and described graphically in a
synoptic way, through the bond-length distributions, which is
presented if figure 2. All graphs in figure 2 correspond to
nanocrystals with the silicon atoms concentrated in the inner core
of the nanoparticle. As we can see the Si-Si distribution has a
peak around 2.48 \AA \ for the first shell of neighbors (connected
to the central atom) and a second peak around the 2.37 \AA for the
rest of the silicon atoms. This second peak, corresponding to
shorter bond-lengths by 0.1 \AA, is more or less constant, with a
tendency to approach the bulk value of 2.36 \AA for larger
nanocrystals. This is also true Ge-Ge bonds. Comparing figures
2(a) and 2(b) we can see that in both cases the bonds of the
central atom with the first shell of neighbors are longer by 0.1
\AA. We also observe in figs. 2(e) and 2(f) that there are no
Ge-Ge bonds, although there is a significant amount of germanium
atoms. Such bonding characteristics are found to be directly
related to both the electronic and optical properties of the
nanoparticles. As was explained earlier, with the same
concentrations ($x$) more than one nanocrystals can be
constructed. Moreover, since the Ge substitutions in the present
work deal with spherical shells of neighbors rather than with
individual atoms, we can distinguish two classes of nanocrystals
with similar concentrations; Those with the Ge atoms in the inner
core, and those with the Ge atoms in the outer shells ("surface").
The structural and cohesive characteristics are different in the
two cases. As we can see in figure 3, we have two distinct curves
depending on the exact location of the Ge layer relatively to the
surface of the nanocrystals. It is clear from this plot that it is
preferable to have the Ge atoms in the "inner" part of the
nanocrystal. This tendency is directly related to the effect of
surface hydrogen atoms and it can be quantified by considering the
binding energy of the independent Si-H($BE_{Si-H}$) and
Ge-H($BE_{Ge-H}$) bonds. This can be approximated by the formulae
\[BE_{Si-H}=\frac{BE_{SiH_4}}{4} \ , \   BE_{Ge-H}=\frac{BE_{GeH_4}}{4} \]
($BE_{SiH_4}$ and $BE_{GeH_4}$ are the corresponding binding
energies of the $SiH_4$ and $GeH_4$ molecules). In this way we can
define the surface energy of the nanoparticle as
\[SE=N_{Si-H}\cdot BE_{Si-H}+N_{Ge-H}\cdot BE_{Ge-H} \]
(where $N_{Si-H}$ and $N_{Ge-H}$ are the number of Si-H and Ge-H
bonds respectively). The dependence of surface energy on the
composition of the nanocrystals and the position of Si and Ge
atoms relative to the surface is shown in figure 3b. It becomes
evident from this figure that the stability of the hydrogenated
clusters is largely determined by their surface. Almost 63\% of
the total binding energy of the nanocrystals is attributed to the
surface Si-H/Ge-H bonds. As a result,the large differences is
surface energy between Ge(core) and Si(core) nanoparticles
(fig.3b) is responsible for the shape and energetic ordering of
the total binding energies shown in fig. 3a. However, it should be
noted that without the hydrogen passivation of the dangling bonds,
it would be natural (energetically favored) for the Ge atoms to
segregate onto the surface in order to minimize the cost of the
dangling bonds. Indeed, as was stated by Tarus et al \cite{19},
for hydrogen-free SiGe nanoclusters, germanium tends to segregate
onto the surface. The above conclusions, are in agreement with
recent theoretical calculation (LDA) of Ramos et. al\cite{21} .
Moreover, the observed trends, are consistent with a series of
experimental data (ree ref,21-26 of Ramos et. al\cite{21})
\subsection{Electronic and optical properties}
\label{opt}

In figure 4 we have plotted the total and  the projected density
of states (DOS and PDOS) for four typical nanocrystals. The DOS
curves were generated from the eigenstates of the ground state
calculations with a suitable gaussian broadening\cite{22}. The
largest variation with the Ge concentration occurs in the valence
band edges, while the conduction band edge is relatively less
sensitive. From these diagrams it can be seen that the hydrogen
contribution lies deep in the valence band (in the energy region
between -10.5 eV to 9.0 eV) leaving Si and/or Ge to dominate the
character of the band edges. For The cases of $Si_5Ge_{42}:H_{60}$
and $Si_{17}Ge_{30}:H_{60}$ nanoparticles (Si in the core and Ge
in the surface) this hydrogen related peak appears to be slightly
broadened. This is probably due to the looser binding of the
hydrogen atoms with the surface Ge atoms (looser as compared with
the corresponding Si-H binding). An interesting conclusion which
can be drawn from the DOS diagrams is related to the character of
the conduction band edges.It appears that when the core (Si or Ge)
is adequately large (i.e. $\sim$ 30\% of the nanoparticle) then
the conduction band edge exhibits a Si or Ge character
respectively. This indicates that excitations of valence electrons
to the Lowest Unoccupied Molecular Orbital (LUMO) or even LUMO+1,
tend to partially confine the excited electron to the nanoparticle
core. Calculations of the HOMO-LUMO gap of mixed $Si_xGe_y:H_z$
nanocrystals have also been recently performed by Yu et al.
\cite{20}, specifically for nanocrystals with a total number of Si
and Ge atoms of 71 (x+y=71). These calculations were based on
density-functional theory (DFT) in the local-density approximation
(LDA). The resulting HOMO-LUMO gaps range from 3.3 - 4.1 eV
corresponding to the pure Ge and pure Si nanocrystals. In order to
compare our calculation with the results of Yu et al.\cite{20} we
performed similar DFT/B3LYP calculations for the pure
$Si_{71}H_{84}$ $Ge_{71}H_{84}$ nanoparticles. Our values of 4.0
eV for the pure Ge nanocrystal and 4.6 eV for the pure Si
nanocrystal are in very good agreement with the values  of Yu et
al. if one considers the inherent tendency of LDA\cite{5} to
underestimate the single particle HOMO-LUMO gap  by approximately
0.6-0.7 eV. A striking difference in the work of Yu et.
al.\cite{20} is that instead of shells of atomic neighbors, used
in the present work (strained nanocrystals), the Ge atoms in ref
\cite{20} are distributed more homogeneously, and they are allowed
to diffuse through the shells (complete relaxation of strain). As
a result the gap dependence on Ge concentration appears to be
practically linear. As a means to provide a more accurate and
detailed account of the optical properties of these nanocrystals
we employed the TDDFT/B3LYP combination in order to calculated
their optical gap (i.e. lowest symmetry and spin allowed
electronic excitation). The results are shown in table 1.The first
comment that can be made by inspecting these values concerns the
nature of the transitions.It is evident that for the nanoparticles
in which the Ge atoms reside in the inner core, the lowest allowed
transition is always between the HOMO- and LUMO orbitals. Moreover
these transition appear to have relatively larger oscillator
strengths. On the contrary, when there is a silicon inner core,
the oscillator strengths are smaller, while the nature of the
transitions becomes more complex. For example, we can see a non
negligible degeneracy concerning the fundamental optical gap
together with an increase on the multireference character of the
transitions. In figure 5 we show a graphic representation  of the
variation of the optical gap as a function of the number of Si
atoms ($x$) contained in the nanocrystal. Both types of
nanoparticles(Si(core) and Ge(core)) are included. We can clearly
distinguish two sets of points (disjoint curves) which  correspond
to the two different types of clusters (Si(core) and Ge(core)). An
analogous variation is also observed for the HOMO-LUMO gap (i.e.
we have an upper and a lower curve). Surprisingly enough, the
larger optical (and HOMO-LUMO) gaps correspond to germanium atoms
lying in the surface region, which as we have seen in figure 3 is
not energetically as stable as the opposite case. Usually, the
most stable structures are the ones which exhibit the largest gap.
However, this rule of thumb seems not to be applicable in this
case. It is interesting to point out that for the case Ge(core)
nanoparticles the gap decreases as the size of the core
increases.This is a common quantum confinement behavior (see for
example ref \cite{4,5,7}). As a result, it may be alleged that the
Ge(core) $Si_xGe_y:H_z$ behave as Ge nanoparticles which are
passivated by a $Si:H$ layer. Comparing the optical gap of
Ge(core) $Si_{42}Ge_5:H_{60}$ nanoparticle with the corresponding
one of $Ge_{5}:H_{12}$ cluster we find that it is smaller by 2.47
eV ( the optical gaps are 3.73 eV for $Si_{42}Ge_5:H_{60}$ and 6.2
eV for $Si_{5}:H_{12}$). This large difference may originate from
a less effective passivation of the Ge core by the Si passivants
(less effective compared to $Ge_5$ passivation by H atoms). To
check this hypothesis we followed a simple line of argument which
goes as follows. By simple calculations on $SiH_4$, $GeH_4$,  and
$H_3Si-GeH_3$ molecules we can find the  binding energies for the
the $Ge-H$, and $Si-Ge$ bonds. In particular, we find (as
expected) that $ BE_{Ge-H} > BE_{Si-Ge}$. Next, we modify the
$Ge-H$ bond length in $GeH_4$ molecule in order to equate the
resulting $BE_{Ge-H}^{*}$ to $BE_{Ge-Si}$. This is achieved when
the $Ge-H$ bond in $GeH_4$ molecule is elongated to $\sim2.1$\AA.
We used this new $Ge^{*}-H$ bond distance for the passivation of
the $Ge_5H_{12}$ cluster and calculated again its optical gap. The
new value is now 3.7 eV and practically coincides with the 3.73 eV
of the Ge(core) $Si_{42}Ge_5:H_{60}$ nanoparticle. This result,
although it does not prove the aforementioned hypothesis, is
highly suggestive of its validity.  The only other
$Si_{x}Ge_{47-x}:H_{60}$ nanoparticle with a Ge core fully capped
(passivated) by Si atoms, suitable for extending the test of our
hypothesis, is $Si_{30}Ge_{17}:H_{60}$ (which should be compared
to $Ge_{17}H_{36}$). However, the $Ge-H$ bond elongation to
$\sim2.1$\AA for the $Ge_{17}H_{36}$ cluster appears problematic
since it brings the hydrogen passivants too close to each other
inducing significant $H-H$ interactions.

On the other hand the variation of the optical gap for the case of
Si(core) $Si_{x}Ge_{47-x}:H_{60}$ nanoparticles as a function of
the Si core size (upper curve in fig.5) appears to be unexpected
(as the Si core increases the gap also increases). This behavior
suggest that the hypothesis of quantum confinement is not
applicable here. In other words, the specific behavior can not be
explained by considering that the Germanium atoms passivate the
inner Si core. Surprisingly, an explanation can be obtained again
by considering the Si atoms of the inner core to passivate
\textquotedblleft internally\textquotedblright the outer shell of
Ge atoms (see fig.6). In order to test this we performed
additional calculation on modified versions of the
$Si_1Ge_{46}:H_{60}$ and $Si_5Ge_{42}:H_{60}$ nanoparticles. In
particular, we removed the inner Si atoms and we passivated the
created internal Ge dangling bonds with hydrogen. The results
indeed show an increase  of the optical gap as we go from
$H_1Ge_{46}:H_{60}$ to $H_5Ge_{42}:H_{60}$. At this point it
should be noted that the \textquotedblleft
internal\textquotedblright  hydrogen passivation is (again) more
effective than the passivation by Si ($BE_{Ge-H} > BE_{Si-Ge}$).
As a result these calculation could only reproduce the trend of
gap increase and not the actual values. Unfortunately, the $Ge-H$
bond elongation to $\sim2.1$\AA which was shown to reproduce the
results of fig. 5 (lower curve) in a quantitative manner can not
be used since it leads to close proximity of adjacent hydrogens in
the interior of the nanoparticle. The conclusions of the last two
paragraphs are summarized in a synoptic way in figure 6. In both
cases (fig6a and fig6b), the optical properties are mainly
determined by Ge part of the nanoparticle while silicon seem to
act mainly as a passivant.

\section{Conclusions}
\label{conc} We have shown that, indeed, the mixed SiGe

nanocrystals have optical and electronic properties intermediate
between those of pure Si and Ge nanocrystals. The large variety of
optical and band gaps depends, not only on the size of the
nanocrystals and the relative concentrations of Si and Ge, but
also on the relative spatial distribution of the Ge atoms with
respect to the surface of the nanocrystals. The stability of the
structures is largely define by the hydrogen surface passivation.
As a result, the most stable nanoparticle are those with the
silicon atoms on the surface (mostly due to the larger binding
energy of the Si-H bonds). The optical properties of Si(core) and
Ge(core) nanoparticles are found to exhibit significant
differences. For the Ge(core) nanocrystals the lowest spin and
symmetry transition are always between the HOMO and LUMO orbitals,
while for the Si(core) ones both HOMO-1 and LUMO+1 contributions
are important (in this case the transitions exhibit a more
pronounced multireference character). The variation of the optical
gap as a function of the core size (Si or Ge) depends drastically
on the nature of the core (Si or Ge). However, for both cases the
optical gap variation can be rationalized by considering that the
silicon atoms behave as simple passivants of the Ge cluster.

These additional degrees of freedom with regard to the properties
of mixed SiGe:H nanoparticles may be important in the future
design of such (and similar) systems, allowing for possible
electronic and optical gap engineering.
\section{Acknowledgments}
\label{ack}
We thank the European Social Fund (ESF), Operational Program for
Educational and Vocational Training II (EPEAEK II), and
particularly the Program PYTHAGORAS, for funding the above work

\clearpage

\begin{table*}[htb]
\begin{center}
\caption{\label{tab1}Lowest Spin and symmetry allowed electronic
transitions}
\begin{tabular}{@{}lllll}
\hline
nanoparticle & core&excitation & Oscillator & Dominant  \\
             &  &energy (eV) &Strength & Contributions \\
\hline
$Si_{46}Ge_1:H_{60}$   &  Ge  & 3.80 & 0.086 & H$\rightarrow$L (98\%)  \\
$Si_{42}Ge_5:H_{60}$   &  Ge  & 3.73 & 0.155 & H$\rightarrow$L (98\%)   \\
$Si_{30}Ge_{17}:H_{60}$&  Ge  & 3.69 & 0.261 & H$\rightarrow$L (97\%)  \\
$Si_{18}Ge_{29}:H_{60}$&  Ge  & 3.54 & 0.201 & H$\rightarrow$L (98\%)  \\
$Si_{12}Ge_{35}:H_{60}$&  Ge  & 3.52 & 0.177 & H$\rightarrow$L (98\%)  \\
$Si_{0}Ge_{47}:H_{60}$ &  Ge  & 3.46 & 0.254 & H$\rightarrow$L (98\%)  \\
$Si_{1}Ge_{46}:H_{60}$ &  Si  & 3.74 & 0.205 & H$\rightarrow$L (97\%)  \\
$Si_{5}Ge_{42}:H_{60}$ &  Si  & 3.88 & 0.017 & H$\rightarrow$L+1 (57\%),  H-1$\rightarrow$ L (35\%) \\
                       &      & 3.92 & 0.110 & H-1$\rightarrow$L (60\%), H$\rightarrow$L+1 (38\%)  \\
$Si_{17}Ge_{30}:H_{60}$&  Si  & 3.91 & 0.067 & H-1$\rightarrow$L (93\%)  \\
                       &      & 4.0  & 0.036 & H$\rightarrow$L+1 (81\%), H-1$\rightarrow$L+1 (17\%)  \\
                       &      & 4.0  & 0.030 & H-1$\rightarrow$L+1 (15\%), H$\rightarrow$L+1 (79\%)  \\
$Si_{29}Ge_{18}:H_{60}$&  Si  & 3.97 & 0.079 & H$\rightarrow$L (96\%)  \\
                       &      & 4.03 & 0.083 & H$\rightarrow$L+1 (97\%)  \\
$Si_{35}Ge_{12}:H_{60}$&  Si  & 3.99 & 0.121 & H$\rightarrow$L (88\%), H$\rightarrow$L+1 (8\%)  \\
$Si_{47}Ge_0:H_{60}$   &  Si  & 4.02 & 0.116 & H$\rightarrow$L (71\%), H$\rightarrow$L+1 (26\%)  \\
\hline
\end{tabular}\\[2pt]
\end{center}
\end{table*}

\clearpage
\begin{list}{}{\leftmargin 2cm \labelwidth 1.5cm \labelsep 0.5cm}
\item[\bf Fig. 1] Typical $Si_xGe_{47-x}:H_{60}$ nanocrystal (a)
$Si_{42}Ge_5:H_{60}$ (b)$Si_{30}Ge_{17}:H_{60}$, (c)
$Si_{5}Ge_{42}:H_{60}$ (d) $Si_{17}Ge_{30}:H_{60}$). The Ge atoms
are shown with green color, while Si atoms are blue. \item[\bf
Fig. 2] Bond distribution in $Si_xGe_{47-x}:H_{60}$ for x=0, 1, 5,
17, 29, 35, 47. The Ge-Ge,  Si-Ge and Si-Si bond distributions are
shown separately. The constant number of the hydrogen atoms (60)
is not shown in the graphs. \item[\bf Fig. 3] (a) Total binding
energy as a function of the number of silicon atoms (b) Surface
energy \item[\bf Fig. 4] Projected and total Density density of
states (PDOS and DOS) of 4 representative nanocrystals. \item[\bf
Fig. 5] The variation of the optical gap as a function of the
number of silicon atoms, for the two categories (Ge(core) and
Si(core)) of $Si_xGe_{47-x};H_{60}$ nanocrystals \item[\bf Fig. 6]
Schematic representation of (a) Ge)core nanoparticle, which
behaves as a Ge nanoparticle passivated by silicon, and (b)
Si(core) nanoparticle which behaves as a hollow Ge nanoparticle
with surface hydrogen passivation and internal Si passivation.
\end{list}
\clearpage
\begin{figure}[ht]
  \begin{center}
   \includegraphics[scale=0.6]{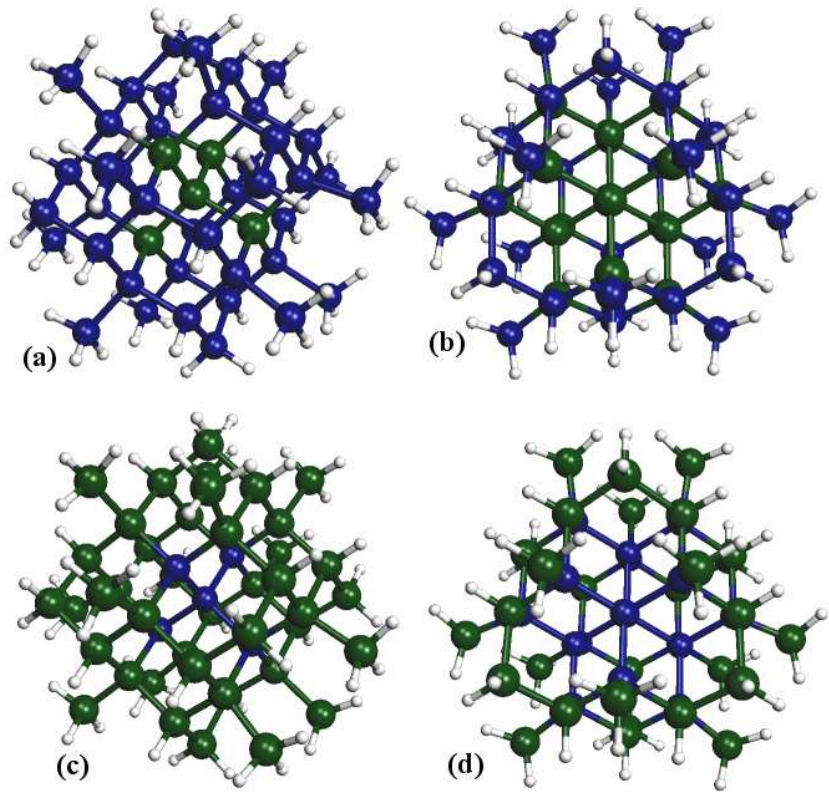}
   \caption{Typical $Si_xGe_{47-x}:H_{60}$ nanocrystal
(a) $Si_{42}Ge_5:H_{60}$ (b)$Si_{30}Ge_{17}:H_{60}$, (c)
$Si_{5}Ge_{42}:H_{60}$ (d) $Si_{17}Ge_{30}:H_{60}$). The Ge atoms
are shown with green color, while Si atoms are blue.}
  \end{center}
\end{figure}
\begin{figure}[ht]
  \begin{center}
   \includegraphics[scale=1.5]{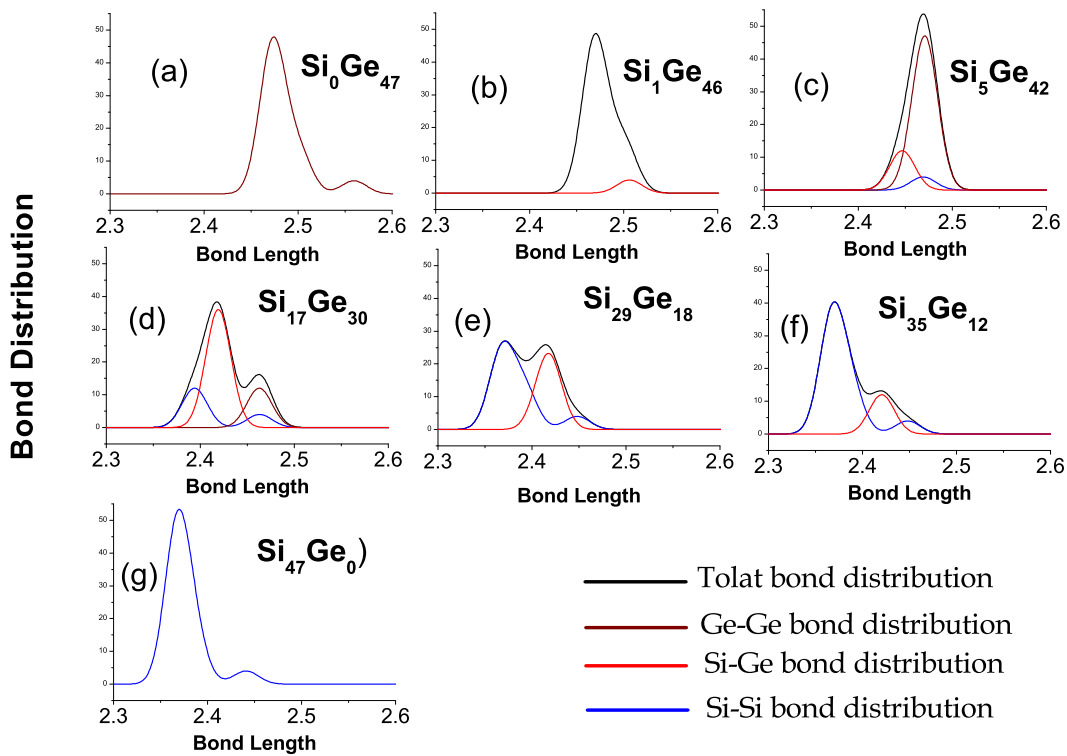}
      \caption{Bond distribution in $Si_xGe_{47-x}:H_{60}$ for x=0, 1,
5, 17, 29, 35, 47. The Ge-Ge,  Si-Ge and Si-Si bond distributions
are shown separately. The constant number of the hydrogen atoms
(60) is not shown in the graphs.}
  \end{center}
\end{figure}
\begin{figure}[ht]
  \begin{center}
     \includegraphics[scale=1.5]{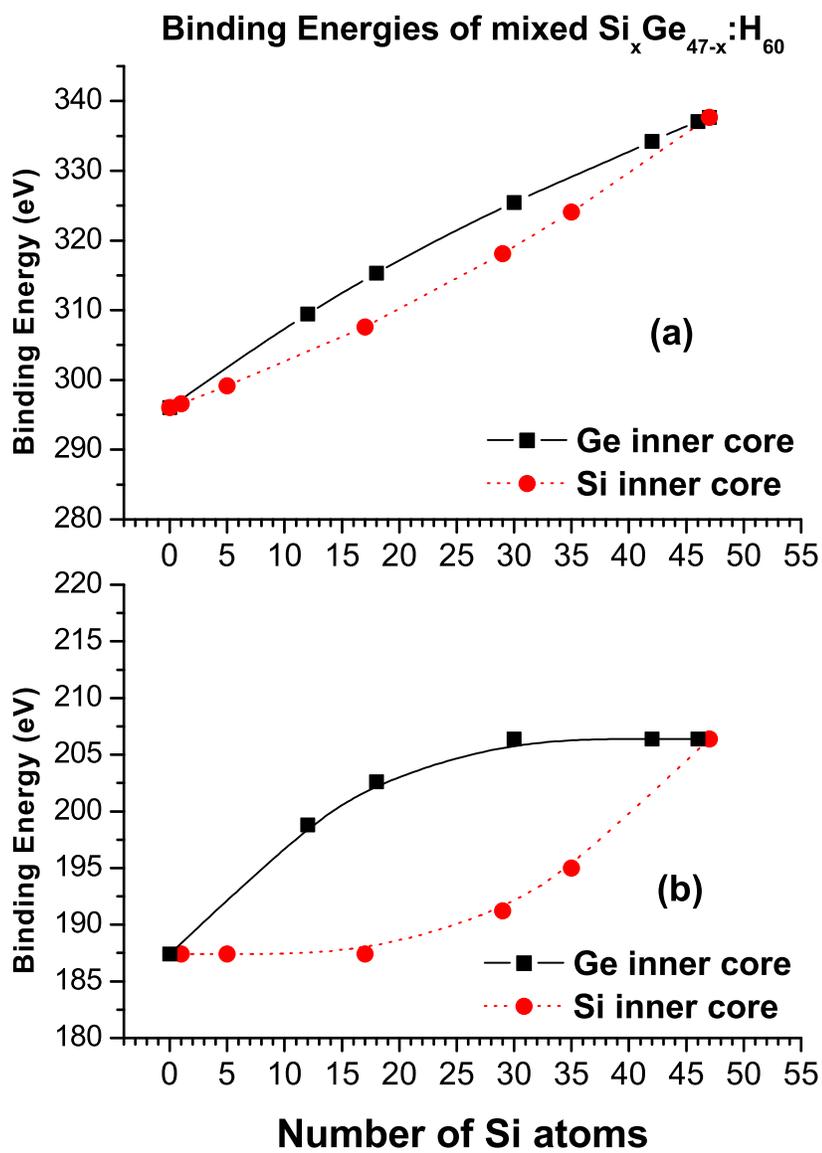}
      \caption{(a) Total binding energy as a function of the number of
silicon atoms (b) Surface energy}
  \end{center}
\end{figure}
\begin{figure}[ht]
  \begin{center}
    \includegraphics[scale=1.0]{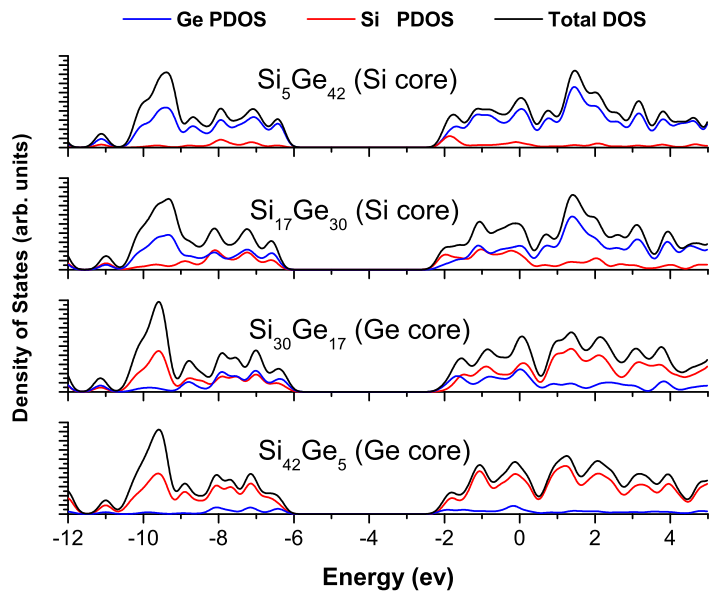}
     \caption{Projected and total Density density of states (PDOS and
DOS) of 4 representative nanocrystals.}
  \end{center}
\end{figure}
\begin{figure}[ht]
  \begin{center}
     \includegraphics[scale=1.0]{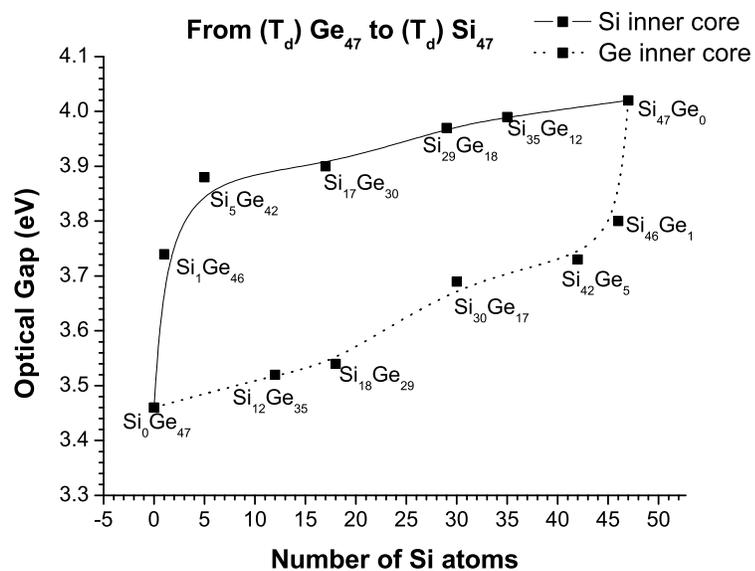}
       \caption{The variation of the optical gap as a function of the
number of silicon atoms, for the two categories (Ge(core) and
Si(core)) of $Si_xGe_{47-x};H_{60}$ nanocrystals}
  \end{center}
\end{figure}

\begin{figure}[ht]
  \begin{center}
     \includegraphics[scale=0.7]{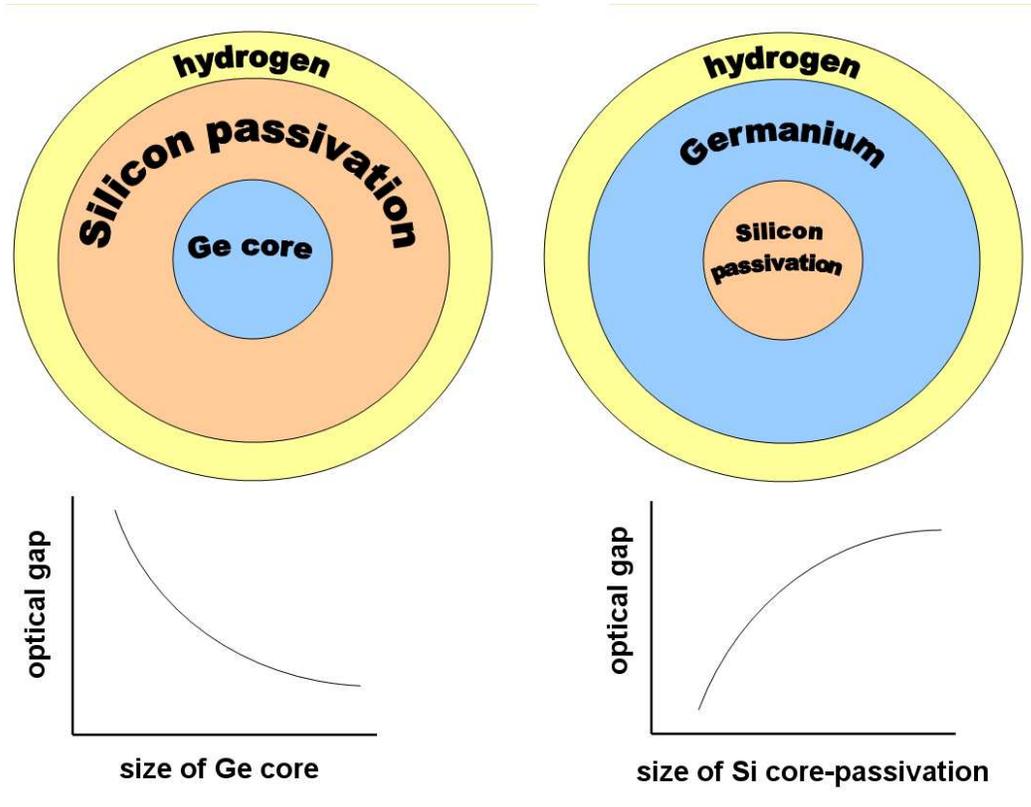}
       \caption{Schematic representation of (a) Ge)core nanoparticle, which behaves as a Ge nanoparticle passivated by silicon, and (b) Si(core) nanoparticle which behaves as a hollow Ge nanoparticle with surface hydrogen passivation and internal Si passivation.}
  \end{center}
\end{figure}

\end{document}